\documentstyle[floats,multicol,aps,epsf,prl]{revtex}

\begin{document}
\draft
\twocolumn[\hsize\textwidth\columnwidth\hsize\csname @twocolumnfalse\endcsname
\title{
Phenomenological Transport Equation for the Cuprate Metals
}
\author{P. Coleman$^1$, A. J. Schofield$^1$ and A. M. Tsvelik$^2$}
\address{
$^1$Serin Laboratory, Rutgers University, P.O. Box 849,
Piscataway, New Jersey 08855-0849}
\address{
$^2$Department of  Physics, University of Oxford, 1 Keble Road,
Oxford OX1 3NP,  UK}
\maketitle
\date{\today}
\maketitle
\begin{abstract}
We observe that the appearance of two transport relaxation times in
the various transport coefficients of cuprate metals may be understood in
terms of scattering processes that discriminate between currents that
are even, or odd under the charge conjugation operator.  We develop a
transport equation that illustrates these ideas and discuss its
experimental and theoretical consequences.
\end{abstract}

\vskip 0.1 truein
\newpage
\vskip2pc]
The normal state of the cuprate superconductors exhibits the
extraordinary feature of two transport relaxation time-scales.  In
optimally doped compounds, conductivity and photo-emission
measurements indicate a scattering rate
which grows linearly with temperature $\tau_{tr}^{-1} =
\eta T$, where for YBCO  $\eta \approx 2$.\cite{linear} By contrast,
Hall constant and magneto-resistance measurements indicate that the
cyclotron relaxation rate $\tau_H^{-1}$ has a qualitatively different
{\em quadratic} temperature dependence.
\begin{eqnarray}
\tau_H^{-1}= \frac{T^2}{W_s} + b_i \; .
\end{eqnarray}
Experimentally, $\tau_H^{-1}$ is inferred from the Hall angle
$\theta_H=\omega_c \tau_H$, manifested in both the Hall conductance
$\sigma_{xy}=\sigma_{xx} \theta_H$ and the magneto-conductance $\Delta
\sigma_{xx}\approx-\sigma_{xx} (\theta_H)^2$.  Experiments on YBCO
demonstrate that $b_i$ is proportional to the impurity concentration
and $W_s$ is estimated to be of order of 800K\cite{chien}. Thus in the
relevant temperature range the ratio of the cyclotron and charge
transport relaxation times $\tau_{tr}/\tau_H \approx T/2W_s$ is small.

This is unprecedented behavior, for in conventional metals, scattering
at the Fermi surface does not discriminate between transverse and
longitudinal currents.  Anderson\cite{phil} has proposed that two
relaxation rates are evidence for two-distinct species of
quasiparticle which independently relax the longitudinal and
transverse currents.\cite{romero}  Two alternative proposals, involving either
strong momentum dependence of the electron
self-energy\cite{carrington,hlubina,pines} or singular
skew-scattering\cite{kotliar} as the origin of two relaxation
time-scales require special conditions to be realized on the Fermi
surface. The former requires that the weakly
scattered parts of the Fermi surface do not short-circuit the
conductivity; the skew scattering model requires near-perfect
particle-hole symmetry.

In this paper we reconsider the idea of two quasiparticle types.  For
its development, this radical idea requires an understanding of how
longitudinal and transverse components of the electro-magnetic current
could couple selectively to two different quasiparticles.  To this
end, we link the discussion with the concept of charge conjugation
symmetry.\cite{majorana,kaons}  Charge-conjugation : the inter-conversion of electrons and
holes, is an asymptotic low-energy symmetry of a Fermi surface.  The
parity $C=\pm 1$ under this symmetry operation delineates longitudinal
electric currents, which are odd ($C=-1$), from transverse currents
and a whole range of other neutral currents, which are even ($C=+1$).
Scattering at the Fermi surface is normally ``blind'' to charge
conjugation symmetry, leading to a single transport relaxation time.
Making the tentative observation that in the cuprates, odd parity
currents relax at the fast rate $\tau_{tr}^{-1}$, whereas other even
parity currents relax at the slow rate $\tau_H^{-1}$, we are led to
hypothesize that new kinds of low energy scattering processes are
present in the cuprate metals which {\em depend on the charge
conjugation symmetry of the quasiparticles}.  By formulating this idea
as a phenomenological transport equation we show that the fastest
relaxation rate dominates the resistivity, but that the slowest
relaxation rate {\em selectively} short-circuits all other current
relaxation processes.  These results constrain a large class
of in-plane thermal and electric transport coefficients, allowing the
hypothesis to be  tested.

Consider a Fermi surface described by the Hamiltonian
\begin{equation}
H_o = \sum_{{\rm \vec p}}
\epsilon_{{\rm \vec p}- e \vec A}\psi^{\dagger}_{{\rm \vec p} \sigma}
\psi_{{\rm \vec p} \sigma} \; .
\end{equation}
We define charge conjugation as
\begin{equation}
\psi_{\rm \vec p \sigma} \longrightarrow
\sigma\psi^{\dagger}_{\rm \vec p^* -\sigma}, \qquad\qquad
\vec A \longrightarrow - \vec A \; ,
\end{equation}
where ${\rm \vec p} = {\rm \vec p}_F + \delta {\rm p} {\rm \hat n}$
and ${\rm \vec p}^{*} = {\rm \vec p}_F - \delta {\rm p} {\rm \hat n}
+O(\delta \rm p^2)$, locate degenerate electron and hole states along
the normal ${\rm \hat n}$ from the Fermi surface.  Physical operators
$\hat O$ can be categorized according to their {\em conserved} parity
under charge conjugation $\hat O \longrightarrow C \hat O$, ($C= \pm
1$).  For example, the electric current operator divides into
independent ``longitudinal'' and ``transverse'' components
$\vec j_e = \vec j_E + \vec j_H$,
\begin{equation}
\begin{array}{rcl}
\displaystyle\vec j_E
&=&\displaystyle e \sum_{{\rm \vec p}} \vec v_{{\rm \vec p}_F}
\displaystyle
\psi^{\dagger}_{{\rm \vec p} \sigma}
\psi_{{\rm \vec p} \sigma} \;\qquad\qquad\qquad\quad
 ( C= -1) ,\nonumber \\
\displaystyle
\vec j_H
&=&\displaystyle
e \sum_{{\rm \vec p}} [
{\underline m}_{\rm \hat p}^{-1} (\delta{ \rm
p}-e \vec A)] \psi^{\dagger}_{{\rm \vec p} \sigma}
\psi_{{\rm \vec p} \sigma} \;\qquad ( C= +1),
\end{array}
\end{equation}
with opposite charge conjugation parities. Here, $\underline{\rm m}_{\rm \hat p}$
is the effective mass tensor.  The transverse current has the the same
$C=+1$ parity as the thermal current operator,  and it is this term
which gives rise to a Hall and thermo-electric response.

Thermal and electric transport is normally described in terms of four
fundamental transport tensors\cite{onsager}
\begin{equation}
\begin{array}{cc}
\vec j_e &= \underline{\sigma} \vec E +  \underline{\beta} \vec
\nabla T  \; , \\
\vec j_t &= \underline{\gamma}\vec E + \underline{\zeta } \vec \nabla
T \; .
\end{array}
\end{equation}
These tensors are directly linked to microscopic charge and thermal
current fluctuations via Kubo formulae.  Table 1. compares the leading
temperature dependences of the various transport tensors measured in
the optimally doped cuprates with a series of calculations we now
describe.  The thermo-electric conductivity $\underline{\beta}$,
determined from the conductivity and Seebeck coefficients, $\underline
S$, $\underline{\beta} = -\underline{ \sigma} \underline{ S}$ has a
particularly revealing temperature dependence.  In a na\"\i ve
relaxation time treatment, the temperature dependence of $\beta$ is
directly related to the relevant quasiparticle relaxation rate
$\tau_{TE}^{-1}$ according to\cite{abrikosov}
\begin{equation}
\beta =  -\left( \pi^2 k_B \over 3 e  \right) \left(
k_B T \over \epsilon_F \right)
{ n e ^2 \over m }\tau_{TE}\; ,
\end{equation}
where $\epsilon_F$ is the Fermi energy.  Combining this with the
electrical conductivity, $\sigma = {ne^2 \over m }\tau_{tr}$, the
dimensionless thermopower is then
\begin{equation}
\tilde S = {eS \over k_B} = \left(
{\tau_{TE} \over \tau_{tr} }  \right)
\left( {\pi^2 \over 3}   \right)
\left( {k_BT \over \epsilon_F}   \right) \; .
\end{equation}
In optimally doped compounds\cite{obertelli},
the thermopower contains an unusual constant part, $\tilde S\approx
\tilde S_o - bT$ where $\tilde S_o\sim 0.1$, which indicates that
\begin{equation}
\tau^{-1}_{TE} = {T^2 / W_{th}} \; ,
\end{equation}
is a factor $T/\eta W_{th}$ smaller than the transport relaxation rate,
where
$W_{th} = ( 3 \tilde S_o/ \pi^2 \eta ) \epsilon_F
\sim {\epsilon_F /10}$.
The comparable size and temperature dependence of $\tau^{-1}_{TE}$ and $ \tau^{-1}_H$
suggest that the same type of quasiparticle carries both the Hall current and the
thermo-current.

\begin{table}
\protect\caption{ Leading temperature dependences of
transport coefficients compared with proposed decomposition into
two Majorana relaxation times (${\cal L}_0$ is the Lorentz number
$\pi^2 k_B^2 / 3 e^2$).}
\begin{center}
\begin{tabular}{ccccc}
Cond-&Majorana&\multicolumn{3}{c}{Leading T behavior}\\
uctivity &Fluid
&$\Gamma_f  \gg \Gamma_s$& & \\
& $\times \left({m \over n e^2} \right)$ &
$(T \gg T^2)$ & Expt. & Ref.\\
\tableline
$\sigma_{\rm xx}$
& ${\displaystyle 2 \over \displaystyle \Gamma_f + \Gamma_s}$
\phantom{$\biggl($}&
$T^{-1}$ &$ T^{-1}$ & \\
$\sigma_{\rm xy}$ &
${\displaystyle \omega_c \over \displaystyle \Gamma_f \Gamma_s}$
\phantom{$\biggl($}
& $T^{-3}$ & $T^{-3}$ \\
$\Delta\sigma_{\rm xx}$ & $-{\displaystyle\sigma_{xx}\over
\displaystyle 2}\bigl(
{\displaystyle {\omega_c^2\over\Gamma_s^2}+{\omega_c^2\over\Gamma_f^2}
}
\bigr)$
\phantom{$\biggl($} & $T^{-5}$ & $T^{-5}$ \\
$\beta_{\rm xx}$ & $-
{\displaystyle eT {\cal L}_0\over \displaystyle 2 \epsilon_F}
\left({\displaystyle 1 \over \displaystyle \Gamma_+}+
{\displaystyle \Gamma_+ \over \displaystyle \Gamma_s\Gamma_f}
\right)
$
& $T^{-1}$ & $T^{-1}$ & \onlinecite{bxx}\\
$ \beta_{\rm xy} $ & $\beta_{\rm xx} {\displaystyle \omega_c \over
\displaystyle \Gamma_+}$ \phantom{$\biggl($}
& $T^{-2}$ & $T^{-3}(?)$ & \onlinecite{bxx} \\
$\zeta_{\rm xx}$ &$ -{{\cal L}_0 \over \displaystyle 2}
\left( {\displaystyle T \over \displaystyle \Gamma_f}+
{\displaystyle T \over \displaystyle \Gamma_s} \right)$
&
$T^{-1}$ & (?) & \onlinecite{kxx} \\
$\zeta_{\rm xy}$ & $ \zeta_{\rm xx} {\displaystyle \omega_c
 \over \displaystyle \Gamma_+}$
\phantom{$\biggl($}&
$T^{-2}$ & $T^{-1}(?)$
& \onlinecite{kxx,ong2} \\
\end{tabular}
\end{center}
\end{table}
By taking linear combinations of degenerate
electron and hole states,
\begin{equation}
\begin{array}{rcl}
\displaystyle
a{}_{{\rm \vec p} \sigma} &=& \displaystyle
{
 {1 \over \sqrt {2}}}
[ \psi_{{\rm \vec p}
\sigma}+ \sigma \psi{^{\dag}}_{{\rm \vec p}^* -\sigma} ],
\qquad(C=+1) \\
\displaystyle
b{}_{{\rm \vec p} \sigma} &=& \displaystyle
{
 {1 \over \mbox{i}\sqrt {2}} }
[ \psi_{{\rm \vec p}
\sigma}- \sigma\psi{^{\dag}}_{{\rm \vec p}^* -\sigma} ],
\qquad(C=-1)
\end{array}
\end{equation}
the low-energy excitations of a Fermi surface described by (1)
may always be re-written as eigenstates of the charge-conjugation
operator\cite{majorana},
\begin{equation}
H_o = \displaystyle\sum_{{\rm |\vec p| > |\vec p_F| }, \sigma}
\Psi{^{\dag}}_{{\rm \vec p} \sigma}
{\rm \epsilon}_{{\rm \vec p -e \vec A \tau_y}}
\Psi_{{\rm \vec p} \sigma} \; .
\end{equation}
where
$
\Psi^{\dagger} _{{\rm \vec p} \sigma} = (
a^{\dagger}_{{\rm \vec p} \sigma},
b^{\dagger}_{{\rm \vec p} \sigma} )$,
and $\tau_y$ is the second Pauli matrix.
Despite the superficial resemblance
with Bogoliubov quasiparticles,
this is merely  an alternative, if unfamiliar
representation of the unpaired  electron gas in terms
of eigenstates of charge-conjugation, rather than eigenstates
of charge.
Note that from (10), photon absorption flips the charge conjugation
parity of the excitation.

In this new basis the Boltzmann f function is a matrix
\begin{equation}
\underline {\rm f}({{\rm \vec p} \sigma}, {\rm \vec R },t)
=
\left[
\begin{array}{cc}
\langle a{^{\dag}}_{{\rm \vec p} \sigma}
a_{{\rm \vec p} \sigma} \rangle  &
\langle b{^{\dag}}_{{\rm \vec p} \sigma} a_{{\rm \vec p} \sigma}
\rangle \\
\langle a{^{\dag}}_{{\rm \vec p} \sigma} b_{{\rm \vec p} \sigma}
\rangle
&   \langle b{^{\dag}}_{{\rm \vec p} \sigma} b_{{\rm \vec p} \sigma}
\rangle
\end{array}
\right]_{\vec {\rm R}, t} \; ,
\end{equation}
where $\langle \ \ \rangle$ represents an appropriate coarse grained
average of the microscopic Green function in the vicinity of $\vec {\rm R}$.
\cite{kadanoff} Electric and thermal currents are given by taking the trace
of $\underline{\rm f}({\rm \vec p})$ with the current operators
\begin{equation}
\displaystyle
j_e({\rm \vec p}) =
\displaystyle e \vec{\cal V}_{{\rm \vec p}} \underline{\tau_y}
\; , \qquad
\displaystyle
\vec j_t({\rm {\rm \vec p}}) = \displaystyle
{\rm \epsilon}_{{\rm \vec p }} \vec {\cal V}_{{\rm \vec p}} \; .
\end{equation}
where $
\vec {\cal V}_{{\rm \vec p}} =
{\vec {\rm v}}_{\rm F}\underline{1} +
\vec {\rm u}_{{\rm \vec p}}\underline{\tau}_y
$
is the velocity operator and  $\vec {\rm u}_{{\rm \vec p}}=
\underline{m}_{\hat {\rm
p}}^{-1}{\rm \delta \vec p }$.
The ``transverse'' current $\vec j_H = e \vec u_{\rm p}$
is diagonal in this basis whereas the ``longitudinal''
current $\vec j_E = e \vec {\rm v}_{\rm F}\tau_y$ is
off-diagonal.
In this representation the Boltzmann equation  becomes
\begin{equation}
\dot {\rm f}  +  {\textstyle{1 \over 2}}
 \{
\vec {\cal V}_{{\rm \vec p}},
\vec \nabla_{\rm R}  {\rm  f} \}
+ {\textstyle{e \over 2}}
 \{ (\vec {\rm E} +
\vec {\cal V}_{{\rm \vec p}}\times \vec{\rm  B} ) \tau_y ,
\vec \nabla_{\rm p} {\rm f} \}
=
%
{\rm I} [ g] \; ,
\end{equation}
where ${\rm I} [g]$ is the collision functional, $g= {\rm f} - {\rm f}
^{(0)}$ is the
departure from equilibrium.
Here the curly brackets
represent anticommutators, which appear when making the gradient
expansion of matrix Green functions.  In this phenomenological
discussion we shall use the relaxation time approximation to the
collision integral, which is
\begin{equation}
{\rm I}[g]= -{1 \over 2} \{ \Gamma , g\} \; ,
\end{equation}
where $\Gamma$ is the relaxation matrix.  For a conventional metal,
where scattering is charge-conjugation invariant,
$\underline{\Gamma} = \Gamma \underline{1}$.

So far, we have merely reformulated conventional transport theory.
Our central phenomenological hypothesis is that {in the cuprate
metals, the relaxation times of the different Majorana modes at the
Fermi surface are no longer equal}. We assign ``fast'' ($\Gamma_f$) and
``slow'' ($\Gamma_s$) scattering rates to quasiparticles of opposite
parity,
\begin{equation}
\underline{\Gamma} = \mbox{diag}[\Gamma_f(T),
\Gamma_s(T)] \; .
\end{equation}
or
$
\underline{\Gamma} =  \Gamma_+ + \Gamma_- \tau_z \; ,
$
where $\Gamma_{\pm} = \frac{1}{2}[\Gamma_f \pm \Gamma_s]$.
Under this assumption, an electron is a linear combination of
 ``fast'' and ``slow''
eigenstates of $\hat C$. Since
$\Gamma_f>> \Gamma_s$, an electron will decay rapidly
in time $\Delta t \sim \Gamma_f^{-1}$ into a quantum admixture of electron and hole:
\begin{equation}
e^-
\stackrel{\quad\Delta t \sim \Gamma_f^{-1}}{\displaystyle
\relbar\mathrel{\mkern-9mu}
\relbar\mathrel{\mkern-9mu}
\relbar\mathrel{\mkern-9mu}
\relbar\mathrel{\mkern-9mu}
\relbar\mathrel{\mkern-9mu}
\relbar\mathrel{\mkern-9mu}
\relbar\mathrel{\mkern-9mu}
\relbar\mathrel{\mkern-9mu}
\relbar\mathrel{\mkern-9mu}
\longrightarrow}
(e^- - h ^ + )/\sqrt{2}
\end{equation}
In this way, charged currents
rapidly decay, leaving behind
a ``neutral'' component which carries the slowly
relaxing Hall, spin, thermal and  thermo-electric currents.
This is an analogue of neutral Kaon decay.\cite{kaons}

Let us now follow these effects in the transport equations.
Writing $g = g_o + \vec {g}\cdot
\vec \tau $  and resolving the components of the transport equation,
we obtain
\begin{equation}
(a + b)  \left(
\begin{array}{cc}
g_0 \\
g_y \\
g_z \end{array} \right)= -
f'\left(
\begin{array}{cc}
\Phi_{{\rm \vec p}} \\
e {\rm \vec E} \cdot \vec{\rm v}_ {\rm F} \\
0 \\
\end{array}
\right) \; ,
\label{theeqn}
\end{equation}
where $f' \equiv \partial_\epsilon f\vert_{\epsilon=
\epsilon_0({\rm \vec p})}$,
$\Phi_{{\rm \vec p}} = e {\rm \vec E} \cdot \vec {\rm u}_{{\rm \vec p}}  -
\epsilon_0({\rm \vec p})T^{-1}\vec \nabla T\cdot \vec {\rm v}_{\rm F}$,
is the neutral current driving term and
\begin{eqnarray}
a &=& \partial_t + \Gamma_+ \underline{1} + \Gamma_-
\underline{\tau}_z \; , \cr
b &=&
\biggl[ e ({\rm \vec E} +  {\cal V}_{{\rm \vec p}} \times {\rm \vec B} )\underline
{\tau}_y
\cdot \vec \nabla_{{\rm \vec p}} +
\vec {\cal V}_{{\rm \vec p}} \cdot \vec \nabla T \partial_T \biggr]
\; ,
\end{eqnarray}
are the collision and gradient terms, in which we have implicitly made
the transformation
\begin{equation}
\textstyle
\underline{\tau}_y \rightarrow
\left(
\begin{array}{ccc}
0 & 1 & 0 \\
1 & 0 & 0 \\
0 & 0 & 0
\end{array}
\right),\qquad
\underline{\tau}_z \rightarrow
\left(
\begin{array}{ccc}
0 & 0 & 1 \\
0 & 0 & 0 \\
1 & 0 & 0
\end{array}
\right).
\end{equation}
The equation for $g_x$ decouples and has been omitted.

To solve the transport equations, we adopt the standard Zener-Jones
multipole expansion, inverting (\ref{theeqn}) and expanding order by
order in powers of $(b/a)$, $g = g^{(1)} + g^{(2)} + \dots$, where
$g^{(n)}= (-a^{-1}b)^{n-1}g^{(1)}$.  By expanding the leading
contributions to the electrical and thermal currents
\begin{equation}
\begin{array}{rcl}
\displaystyle
\vec j_e &=&\displaystyle
 e \sum \bigl[
{\rm \vec v}_{\rm F} g_{y}({\rm \vec p}) +
{\rm \vec u}_{\rm \vec p} g_0({\rm \vec p})\bigr] \; ,\\
\displaystyle
\vec j_t &=&\displaystyle
  \sum
\epsilon({\rm \vec p})
\bigl[
{\rm \vec v}_{\rm F}
g_0({\rm \vec p}) +
{\rm \vec u}_{\rm \vec p}
g_y({\rm \vec p})\bigr]
 \; .
\end{array}
\end{equation}
\begin{figure}[btp]
\epsfxsize=3.0in
\epsfbox{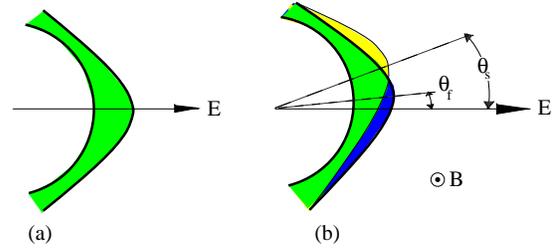}
\protect\caption{(a) Application of field creates a mixture of
slowly and rapidly relaxing quasiparticles. (b) Slow and fast component
of the Majorana fluid precesses in a field, equilibrating
at large and small Hall angles respectively.
}
\label{Fig1}
\end{figure}
\noindent
we obtain the transport coefficients. The results for a simplified
parabolic band are summarized  in Table 1.

A simple physical picture of the effect of an electric field is
provided in Fig. 1.  When an electric field is applied, it produces an
admixture of $C=+1$ and $C=-1$ quasiparticles whose joint relaxation
rate $\Gamma_{tr}=\frac{1}{2}[\Gamma_s + \Gamma_f]\approx
\frac{1}{2}\Gamma_f$ is dominated by the rapidly  relaxing
quasiparticles.  Magnetic fields couple diagonally to the Majorana
quasi-particles, deflecting each component through a
Hall angle $\theta_{s,f} = \omega_c/ \Gamma_{s,f}$.
Since $\theta_s>>\theta_f$. the Hall current is dominated by the
slow-relaxation quasiparticles.

A thermal gradient couples diagonally to the quasiparticles, so
thermal and thermo-electric conductivities are determined by the slow
relaxation rate.  The difference in the relaxation times of the
electrical and thermo-electric currents then gives rise to the unique
temperature independent component in the Seebeck coefficient $S =
-\rho \beta \propto (T\Gamma_f / \Gamma_s)$.
The off-diagonal field-dependent part of the thermal conductivity, is
of interest because it is free from phonon contributions.  The
field-dependent part of thermal current is even under the charge
conjugation operator, so the thermal Hall angle is determined by the
fast relaxation rate, $\theta_T \sim \omega_c/\Gamma_+$, giving
$\zeta_{xy} \sim 1/T^2$. Provisional measurements of the Hall
conductivity\cite{ong2} show that it grows as the temperature is
lowered, but suggest $\zeta_{xy} \sim 1/T$, a result which if
sustained, would refute our approach.

Various experiments can be used both to test and contrast our picture
with alternative theories. Most importantly, we predict that the fast
relaxation rate will only appear in charge-conjugation-odd currents;
all other currents will be short-circuited by the slowly
relaxing quasiparticles.
A.C. Hall conductivity $\sigma_{xy}(\omega)$
is another discriminatory probe.  Provided $\Gamma_s<<\Gamma_f$,
our model predicts
\begin{equation}
\cot \theta_H(\omega)={\sigma_{xx}(\omega)\over \sigma_{xy}(\omega)} =
\left( \frac{-i\omega + \Gamma_s(T)}{\omega_c}\right) \; . \label{hallangle}
\end{equation}
In the skew-scattering model\cite{kotliar},
$\omega_c\rightarrow\omega_c^*(T)\propto{1 \over T}$ is renormalized and
there is only one relaxation rate $\Gamma_s= \Gamma_f$, so $Im[\cot
\theta(\omega)] \propto \omega T $ is proportional to temperature. By
contrast, in the two-relaxation time scenario, this quantity is
temperature independent.  The extension of existing
A.C. Hall measurements on $YBCO$\cite{Drew} to a variety of
temperatures can thus delineate these two scenarios.

Symmetry arguments  have led us to suggest
that if Hall and electric currents decay at qualitatively different
rates, the imaginary part of the electron self-energy  has the
form
\begin{equation}
\Gamma = \Gamma_+ + \Gamma_- \hat C,
\end{equation}
where $\hat C$ is the charge conjugation operator.
We now discuss the interpretation of this hypothesis.
Microscopically, terms
proportional to $\hat C$  represent the inelastic
inter-conversion of electrons and holes.  This is superficially
similar to ``charge imbalance relaxation'' in
superconductors, where inelastic collisions give rise to a relaxation
of quasiparticle charge into the condensate at a rate
$\Gamma_Q$.\cite{clarke,charge}
Evidently we have no condensate,
but for consistency we do  need
a coherent charge reservoir to which charge
is transferred by anomalous scattering events; its coherence length
has to be {\em finite}, but it must also be
long compared with
the quasiparticle mean-free path
\begin{equation}
\xi(T) \gtrsim  { v_F /
\Gamma_s(T)} \; .
\end{equation}
We are thus tempted to
interpret $\Gamma_-\sim 2 T$ as a charge relaxation rate which
survives in the normal state by virtue of these large coherent
patches.  It may be possible to directly test this idea, for
once the cuprate metals become
superconducting, the charge relaxation rate can be measured using a
NSN tunnel junction\cite{clarke} or phase slip centers in narrow
wires.   In conventional superconductors as
$T\rightarrow T_c$, $\Gamma_Q\rightarrow 0$; \cite{clarke,charge}
in the cuprates we expect the charge imbalance decay rate
to remain finite as  $T\rightarrow T_c$.

These  lines of reasoning suggest
the presence of a mutual
inelastic decay channel for electrons and holes $ h^+ \rightleftharpoons
\ {\rm neutral\ state} \rightleftharpoons e^-$
which involves
the emission or
absorption of a charge $+1e$ object into the charge reservoir.
Perhaps there is a loose link here with the
``holon'' excitation in
Anderson's Luttinger liquid scenario?\cite{phil}
We  also note that
recent theoretical work on non-Fermi liquid behavior in
impurity models indicates one origin of
scale-invariant marginal Fermi liquid behavior\cite{margin}
is the formation of fermionic three-body bound-states with
{\em definite} charge conjugation symmetry.\cite{varma,poor}
The formation of such objects around the Fermi surface,
might provide a microscopic basis for our phenomenological hypothesis.

In summary, we have proposed that the appearance of two transport
relaxation times in the cuprate metals is a consequence of
scattering effects that are sensitive to the charge
conjugation symmetry of the quasiparticles.  We have formulated this
hypothesis in a transport equation where one
eigenstate of charge-conjugation symmetry decays much more rapidly
than its partner. The decay of  electric
current is dominated by the rapid relaxation rate while ``neutral''
transport currents, including the Hall and thermo-electric
currents, are governed by the slower decay rate.

We should like to thank E. Abrahams, P. W. Anderson,
G. L. Lonzarich and Ph. Ong for discussions related to this work.
We are indebted to O. Entin-Wohlman  for identifying a
possible link with charge transfer imbalance in superconductors.
This research was supported by
travel funds from NATO grant CRG 940040 (AMT \& PC), NSF grants
DMR-93-12138 (PC) and a travel fellowship from the Royal Society
(AJS).


\begin{references}
\bibitem{onsager} $\underline{\beta}$ and
$\underline{\gamma}$ are linked by the Onsager
relation  $\underline{\gamma}(B) = - T \underline{\beta}^T(-B) $, where
$B$ is the magnetic field.
\end{references}
\end{document}